\documentclass[AMA,Times1COL]{WileyNJDv5}

\articletype{Research Article}

\received{Date Month Year}
\revised{Date Month Year}
\accepted{Date Month Year}
\journal{Journal}
\volume{00}
\copyyear{2024}
\startpage{1}

\usepackage{amsmath}
\usepackage{graphicx}
\usepackage{xr}
\usepackage{bm}
\def\bSig\mathbf{\Sigma}

\newcommand{\bi}{\begin{itemize}}
\newcommand{\ei}{\end{itemize}}
\newcommand{\0}{\bm{0}}
\newcommand{\Y}{\bm{Y}}

\newcommand{\y}{\bm{y}}

\newcommand{\bY}{\bm{Y}}

\newcommand{\x}{\bm{x}}
\newcommand{\btheta}{\bm{\theta}}
\newcommand{\bmu}{\bm{\mu}}
\newcommand{\bSigma}{\bm{\Sigma}}

\newcommand{\bDelta}{\bm{\Delta}}

\newcommand{\bx}{\bm{x}}
\newcommand{\bz}{\bm{z}}

\newcommand{\bb}{\bm{b}}

\newcommand{\A}{\mbox{\textbf{A}}}

\newcommand{\Var}{\mbox{Var}}

\newcommand{\Norm}{\mathcal {N}}

\begin{document}

\title{A Reflection on the Impact of Misspecifying Unidentifiable Causal Inference Models in Surrogate Endpoint Evaluation}

\author[1]{Gokce Deliorman}

\author[2]{Florian Stijven}

\author[3]{Wim Van der Elst}

\author[1,4]{Maria del Carmen Pardo}

\author[2]{Ariel Alonso}

\authormark{AUTHOR ONE \textsc{et al}}

\titlemark{A Reflection on the Impact of Misspecifying Unidentifiable Causal Inference Models in Surrogate Endpoint Evaluation}

\address[1]{\orgdiv{Department of Statistics and O.R}, \orgname{Complutense University of Madrid}, \orgaddress{\state{Madrid}, \country{Spain}}}

\address[2]{\orgdiv{L-BioStat}, \orgname{KU Leuven}, \orgaddress{\state{Leuven}, \country{Belgium}}}

\address[3]{\orgdiv{Janssen Pharmaceutica}, \orgname{Companies of Johnson \& Johnson}, \orgaddress{\state{Leuven}, \country{Belgium}}}

\address[4]{\orgdiv{Interdisciplinary Mathematics Institute (IMI)}, \orgname{Complutense University of Madrid}, \orgaddress{\state{Madrid}, \country{Spain}}}

\corres{Gokce Deliorman, Faculty of Mathematics,
Plaza Ciencias 3,
Complutense University of Madrid, 
28040 Madrid (Spain).
\email{gdeliorm@ucm.es}} 

\fundingInfo{Spanish Ministry of Science and Innovation, Grant/Award
Number:PID2022-137050NB-I00,\\
Agentschap Innoveren \& Ondernemen and Janssen Pharmaceutical Companies of Johnson \& Johnson Innovative Medicine through a Baekeland Mandate, Grant/Award
Number:HBC.2022.0145}

\abstract[Abstract]{Surrogate endpoints are often used in place of expensive, delayed, or rare true endpoints in clinical trials. However, regulatory authorities require thorough evaluation to accept these surrogate endpoints as reliable substitutes. One evaluation approach is the information-theoretic causal inference framework, which quantifies surrogacy using the individual causal association (ICA). Like most causal inference methods, this approach relies on models that are only partially identifiable. For continuous outcomes, a normal model is often used. Based on theoretical elements and a Monte Carlo procedure we studied the impact of model misspecification across two scenarios: 1) the true model is based on a multivariate t-distribution, and 2) the true model is based on a multivariate log-normal distribution. In the first scenario, the misspecification has a negligible impact on the results, while in the second, it has a significant impact when the misspecification is detectable using the observed data. Finally, we analyzed two data sets using the normal model and several D-vine copula models that were indistinguishable from the normal model based on the data at hand. We observed that the results may vary when different models are used.}


\jnlcitation{\cname{
\author{Deliorman G.},
\author{Stijven F.},
\author{Van der Elst W.},
\author{Pardo M.C.}, and
\author{Alonso A.}}.
\ctitle{A Reflection on the Impact of Misspecifying Unidentifiable Causal Inference Models in Surrogate Endpoint Evaluation.} \cjournal{\it Statistics in Medicine.} \cvol{2024;00(00):1--18}.}

\maketitle

\renewcommand\thefootnote{}

\renewcommand\thefootnote{\fnsymbol{footnote}}
\setcounter{footnote}{1}

\section{Introduction}
\label{sec:intro}
Both humanitarian and commercial considerations have sparked a relentless search for methods to expedite the development of new therapies while also reducing the associated costs. In response to this pressing need, researchers and medical professionals have eagerly explored the potential of surrogate endpoints, a general strategy that has gathered considerable interest over the last decades. Surrogate endpoints are alternative measures that can either replace or supplement the most clinically relevant outcome, the so-called true endpoint, when evaluating experimental treatments or interventions.

The main advantage of surrogate endpoints lies in their potential to be measured earlier, more conveniently, or more frequently than the true endpoint. This advantage not only accelerates the pace of clinical trials, but also streamlines the evaluation process, making it more efficient and resource-friendly. As a result, regulatory agencies across the globe, specially those in the United States, Europe, and Japan, began to introduce policies to regulate the use of surrogate endpoints in the evaluation of new treatments \citep{FDA2014,Burzykowski2005}.

However, the critical question is how to establish the adequacy of a surrogate endpoint. In other words, how can we ensure that the treatment's effect on the surrogate accurately predicts its impact on a more clinically meaningful true endpoint. To tackle this challenge, researchers have proposed various definitions of validity and formal criteria. These approaches can be categorized into two frameworks: methodologies relying on single trial data (single trial setting or STS) and methodologies using data from multiple clinical trials. The meta-analytic framework, based on expected causal treatment effects across multiple clinical trials, is considered one of the most general and effective methods for evaluating surrogate endpoints \citep{weir2022informed}. However, its practical implementation is hindered by stringent data requirements, which are often unavailable in the early stages of drug development when surrogate endpoints are most needed \citep{Burzykowski2005,Alonso2017,FDA2014}. As a result, developing new methods for the STS remains a critical goal in the field. Methods developed in the STS frequently rely on individual causal treatment effects and assess the surrogate's validity in a fixed and well-defined population.

In the present work, the information-theoretic causal inference (ITCI) framework is considered in a single-trial setting. In this scenario, Alonso (2018) \cite{alonso2018} introduced a general definition of surrogacy based on the concept of information gain or, equivalently, uncertainty reduction. To assess the definition a general metric of surrogacy, the so-called individual causal association (ICA), was proposed. The ICA quantifies the association between the individual causal treatment effects on the surrogate and true endpoint based on their mutual information. Additionally, Deliorman et al. \cite{deliorman2024} extended the ICA using the Rényi divergence, with the original definition being a special case of this broader formulation.

Assessing the ICA requires a joint causal inference model for the potential outcomes of both variables and a suitable re-scaling of the mutual information that satisfies desirable mathematical properties. When both outcomes are continuous, Alonso et al. (2015) \cite{alonso2015relationship} introduced a causal inference model based on a four-dimensional normal distribution and proposed an appropriate quantification of the ICA. The model is only partially identifiable, and hence, the ICA cannot be estimated from the data without making strong unverifiable assumptions. This issue has been addressed through a simulation-based sensitivity analysis for the normal causal model. The unidentifiability of the model also raises concerns about the impact that potential misspecifications may have on the ICA value and, consequently, on the surrogate's validity assessment. To our knowledge, this problem has not been thoroughly investigated, despite its significant practical implications. In this work, we tackle it using both theoretical elements and Monte Carlo methods.

The structure of this paper is as follows: In Section \ref{sec:ICA-MtD}, we introduce a general framework for evaluating surrogacy when both outcomes are continuous in a single-trial setting. Section \ref{sec:Normal} revisits the normal causal model. In Section \ref{sec:multi_t_dist}, we examine the impact of misspecification when the true data-generating mechanism is a multivariate t-distribution, using some theoretical elements. Section \ref{sec:ext-alg} extends the sensitivity analysis algorithm initially proposed by Alonso et al. (2015) \cite{alonso2015relationship} to accommodate non-normal models. Subsequently, in Section \ref{sec:lognormal}, we investigate the effect of misspecification when the true data-generating mechanism follows a multivariate log-normal distribution using the extended algorithm. In Section \ref{sec:copula}, we analyze two case studies involving multiple causal models, including the normal causal model and several D-Vine copula models, all of which are indistinguishable based on the available data. Finally, Section \ref{sec:discuss} provides concluding remarks.

\vspace{-5mm}
\section{Assessing Surrogacy with Continuous Outcomes}
\label{sec:ICA-MtD}

In the following, we summarize the methodology introduced by Alonso et al. (2015) \cite{alonso2015relationship} and Alonso (2018)\cite{alonso2018} for evaluating the validity of a proposed continuous surrogate endpoint $S$ for a continuous true endpoint $T$. This evaluation is conducted using data from a single randomized clinical trial with a well-defined population, where only two treatments ($Z=0/1$) are being assessed in a parallel study design.

The potential outcomes model of Neyman and Rubin assumes that each patient has a four dimensional vector of potential outcomes $ \bY = (T_0,T_1,S_0,S_1)^T$ with $S_z$ and $T_z$ representing the outcome for the surrogate and true endpoint under treatment $Z=z$. The practical implementation of the model is based on several assumptions \citep{imbens2015causal}. 
First, the stable unit treatment value assumption (SUTVA) links the observed outcomes to the potential outcomes as follows: 
$${(S, T)^T = Z \cdot (S_1, T_1)^T} {+ (1 - Z) \cdot (S_0, T_0)^T}.$$
Second, the full exchangeability assumption states that the potential outcomes are independent of the assigned treatment: $(T_0,T_1,S_0,S_1)^T \perp Z$.  These two assumptions are typically met in randomized clinical trials and they will be used throughout the remainder of this paper.

The vector of individual causal treatment effects is defined as $\bDelta= (\Delta T, \Delta S)^T$ where $\Delta S = S_1 - S_0$ and \mbox{$\Delta T = T_1 - T_0$}. Alonso (2018) \cite{alonso2018} introduced the following definition of surrogacy in the STS.
\begin{definition}
\label{def:surrogate}
In the STS, we shall say that $S$ is a good surrogate for $T$ if $\Delta S$ conveys a substantial amount of information on $\Delta T$.
\end{definition}
The concept of information has been rigorously defined in information theory \citep{cover1991information}. The amount of ``shared" information between $\Delta S$ and $\Delta T$ can be quantified using the mutual information between these individual causal treatment effects, denoted by $I(\Delta T, \Delta S)$. Mutual information is always non-negative, zero if and only if $\Delta S$ and $\Delta T$ are independent, symmetric, and invariant under bijective transformations. Despite its appealing mathematical properties, interpreting mutual information can be challenging because it lacks an upper bound when $\Delta S$ and $\Delta T$ are continuous. This issue is addressed by mapping mutual information onto the unit interval, ensuring it takes a value of zero when $\Delta T$ and $\Delta S$ are independent and one when there is a non-trivial transformation $\phi$ such that $\Delta T=\phi(\Delta S)$ with probability one. 

The mutual information between $\Delta S$ and $\Delta T$ is a functional of the joint distribution $f(\Delta T, \Delta S)$, which is completely determined by the distribution of the vector of potential outcomes $\Y$. In the following sections, several parametric models for this distribution will be proposed. We start with the widely used multivariate normal causal model, which serves as the reference model, and later consider other models such as the multivariate t and log-normal causal models.

\section{The Normal Causal Model}
\label{sec:Normal}

Alonso et al. (2015) \cite{alonso2015relationship} assumed that $\Y\sim \Norm(\bmu,\bSigma)$ with $\bmu=(\mu_{T0},\mu_{T1},\mu_{S0},\mu_{S1})^T$ and
\begin{equation}
\label{eq:Varcounter}
\bSigma=\left(\begin{array}{cccc}
\sigma_{T0T0}&\sigma_{T0T1}&\sigma_{T0S0}&\sigma_{T0S1}\\
\sigma_{T0T1}&\sigma_{T1T1}&\sigma_{T1S0}&\sigma_{T1S1}\\
\sigma_{T0S0}&\sigma_{T1S0}&\sigma_{S0S0}&\sigma_{S0S1}\\
\sigma_{T0S1}&\sigma_{T1S1}&\sigma_{S0S1}&\sigma_{S1S1}
\end{array}\right).
\end{equation}
Under this assumption, $\bDelta=\left(\Delta T, \Delta S\right)^T=\A\Y\sim \Norm \left(\bmu_{\Delta},\bSigma_{\Delta}\right)$, where $\bmu_{\Delta}=(\beta,\alpha)^T$, $\beta=E(\Delta T)$, $\alpha=E(\Delta S)$ and $\bSigma_{\Delta}=\A\bSigma\A^T$ with $\A$ the corresponding contrast matrix. Furthermore, these authors proposed to assess Definition \ref{def:surrogate} using the Squared Information Correlation Coefficient (SICC) \citep{linfoot1957informational, joe1989relative}
\begin{equation}
\label{eq:R2H}
 R_H^2=1-{\displaystyle e^{-2I(\Delta T, \Delta S)}}
\end{equation}
where $I(\Delta T, \Delta S)$ is the mutual information between $\Delta T$ and $\Delta S$.
For continuous outcomes the SICC satisfies the properties given at the end of Section \ref{sec:ICA-MtD} and, therefore, one may argue that (\ref{eq:R2H}) is a suitable metric to assess Definition \ref{def:surrogate}, i.e., one may argue that it is a good metric of surrogacy. Alonso et al. (2015) \cite{alonso2015relationship} called this metric the individual causal association or ICA. Under the normality assumption, $-2I(\Delta T, \Delta S)=\log(1-\rho_{\Delta}^2)$ where $\rho_{\Delta}=\mbox{corr}\left(\Delta T,\Delta S\right)$ and, consequently, $R_H^2=\rho_{\Delta}^2=ICA_N$. Based on this equivalence, Alonso et al. (2015) \cite{alonso2015relationship} proposed to quantify the ICA using the Pearson's correlation coefficient between the individual causal treatment effects. Another important reason behind this choice is that correlations are more widely known and better understood by practicing clinicians than the SICC. In addition, it can be shown that
$$
\rho_{\Delta}=\dfrac{\sqrt{\sigma_{T0T0}\sigma_{S0S0}}\rho_{T0S0}+\sqrt{\sigma_{T1T1}\sigma_{S1S1}}\rho_{T1S1}-\sqrt{\sigma_{T1T1}\sigma_{S0S0}}\rho_{T1S0}-\sqrt{\sigma_{T0T0}\sigma_{S1S1}}\rho_{T0S1}}
{\sqrt{\left(\sigma_{T0T0}+\sigma_{T1T1}-2\sqrt{\sigma_{T0T0}\sigma_{T1T1}}\rho_{T0T1}\right)\left(\sigma_{S0S0}+\sigma_{S1S1}-2\sqrt{\sigma_{S0S0}\sigma_{S1S1}}\rho_{S0S1}\right)}},
$$
where $\rho_{XY}$ denotes the correlation between the potential outcomes $X$ and $Y$. The ICA is also a measure of prediction accuracy, i.e., a measure of how accurately one can predict the causal treatment effect on the true endpoint for a given individual, using his causal treatment effect on the surrogate. If one further makes the homoscedasticity  assumptions $\sigma_{T0T0}=\sigma_{T1T1}=\sigma_{T}$ and $\sigma_{S0S0}=\sigma_{S1S1}=\sigma_{S}$, i.e., the variability of the true and surrogate endpoint is constant across the two treatment conditions, then $\rho_{\Delta}$ takes the simpler form
$$
\rho_{\Delta}=\dfrac{\rho_{T0S0}+\rho_{T1S1}-\rho_{T1S0}-\rho_{T0S1}}{2\sqrt{\left(1-\rho_{T0T1}\right)\left(1-\rho_{S0S1}\right)}}.
$$
Some comments come in place. The so-called fundamental problem of causal inference states that, in practice, only two of these four potential outcomes are observed, and, hence, the distribution of $\Y$ is not identifiable \citep{holland1986statistics}. Therefore, the vector of potential outcomes $\Y$ is essentially unobservable. Firstly, note that although $\rho_{T0S0}$ and $\rho_{T1S1}$ are identifiable from the data, the other correlations are not and, as a result, the ICA cannot be estimated without imposing untestable restrictions on the unidentifiable correlations. Secondly, the previous expressions clearly illustrate that assumptions about the association between the potential outcomes for the surrogate ($\rho_{S0S1}$) and the true endpoint ($\rho_{T0T1}$) may have an impact on ICA and, consequently, on the assessment of surrogacy. Alonso et al. (2015) \cite{alonso2015relationship} addressed this problem by considering the four-dimensional subset 
$$\Gamma_D=\left\{\btheta=\left(\rho_{T1S0},\rho_{T0S1},\rho_{T0T1},\rho_{S0S1}\right): \mbox{ so that } \widehat {\bSigma} \mbox { is positive definite} \right\}$$ 
with $\widehat{\bSigma}$  denoting the matrix $\bSigma$ with the identifiable entries substituted by their estimated values. The ICA is a mathematical function of the unidentifiable correlations in $\Gamma_D$. In theory, one could study the behavior of $\rho_{\Delta}(\btheta)$ on $\Gamma_D$ through a purely mathematical approach. However, this method presents significant challenges. A more pragmatic solution is to tackle this problem using a stochastic procedure by sampling a sufficiently large number of $\btheta$ vectors in $\Gamma_D$ \citep{alonso2015relationship}. For each element of this sample, the joint distribution of $\Y$ can be fully determined, allowing the calculation of the joint distribution of the individual causal treatment effects $\bDelta = (\Delta T, \Delta S)^T$ and the corresponding ICA. The frequency distribution of the resulting $\rho_{\Delta}(\btheta)$ values would then provide insights into its behavior on $\Gamma_D$, and consequently, the validity of the surrogate across all scenarios compatible with the data.

\section{The Multivariate t-distribution Causal Model}
\label{sec:multi_t_dist}
When the surrogate and true endpoints are continuous outcomes, the ICA was quantifed under the assumption that the vector of potential outcomes $\Y$ followed a multivariate normal distribution. This normality assumption played an important role in the deduction of the ICA theoretical properties presented in the literature as well as its implementation in the R package Surrogate \citep{VanderElst2016,VanderElst2021,Alonso2019}. This raises the question about the sensitivity of $\rho_{\Delta}$ to departures from the normal model. In this section, this issue is studied using a multivariate t-distribution. 

\subsection{Theoretical Background}
\label{sec:MtD}

The multivariate t-distribution is a special case of a more general family, the so-called elliptical distributions \citep{arellano2013shannon}. One way of defining a p-dimensional multivariate t-distribution is based on the fact that if $\y \sim \Norm(\0, \bSigma)$ and $u \sim \chi_{\nu}^2$ with $\y \perp u$ then $\x= \bmu+\y/\sqrt{u+\nu}$ follows a multivariate t-distribution with density function
\begin{equation*}
   f(\x|\bmu, \bSigma, \nu)=\frac{\Gamma[(\nu+p)/2]}{\Gamma(\nu/2)(\nu \pi)^{p/2} |\bSigma|^{1/2} } \Big [ 1+ \frac{1}{\nu}(\x-\bmu)^T \bSigma^{-1}(\x-\bmu) \Big ]^{-(\nu+p)/2}
\end{equation*}
The multivariate t-distribution has parameters $\bSigma$, $\bmu$, $\nu$ and it is denoted as $\x \sim t_p(\bmu, \bSigma, \nu)$. The expected value of $\x$ equals $E(\x) = \bmu $ for $ \nu > 1$ (else undefined), however, $\bSigma$ is not the covariance matrix of $\x$ since Var$(\x)=\nu /(\nu-2)\bSigma$ for $(\nu>2)$. An important special case is obtained when $\nu=1$ that is the so-called multivariate Cauchy distribution. The multivariate t-distribution has some interesting mathematical properties. For instance, let us consider the $p$-dimensional vector
\begin{equation}
\label{eq:dec1}
\x=\begin{pmatrix}\x_1\\ \x_2\end{pmatrix}\sim t_p(\bmu,\bSigma,\nu)
\end{equation}
with $\bx_i$ a $p_i$-dimensional vector ($p_1+p_2=p$) and let us further consider the partitions
\begin{equation}
\label{eq:dec2}
\bmu=\begin{pmatrix}\bmu_1\\\bmu_2\end{pmatrix} \quad\mbox{and}\quad \bSigma=\begin{pmatrix}\bSigma_{11} & \bSigma_{12}\\\bSigma_{21} & \bSigma_{22}\end{pmatrix}.
\end{equation}
\cite{ding2016conditional} shows that 
$$\x_2\mid\x_1\sim t_{p_2}\left(\bmu_{2|1},\dfrac{\nu+d_1}{\nu+p_1}\bSigma_{22|1},\nu+p_1\right)$$
with the following conditional mean $\bmu_{2|1}=\bmu_{2}+\bSigma_{21}\bSigma_{11}^{-1}\left(\x_1-\bmu_1\right)$, conditional variance $\bSigma_{22|1}=\bSigma_{22}-\bSigma_{12}\bSigma_{11}^{-1}\bSigma _{21}$ and $d_1=(\x_1-\bmu_1)^T\bSigma_{11}^{-1}(\x_1-\bmu_1)$.

The multivariate t-distribution is particularly appealing for our purposes because, if the vector of potential outcomes follows this distribution, then the patient-level treatment effects will also follow a multivariate t-distribution and an analytical expression for the mutual information is available in this context. Kotz and Nadarajah (2004)\cite{kotz2004multivariate} present an important result regarding the distribution of a linear combination of a $p$-dimensional t-distributed vector. In fact, these authors show that if $\x\sim t_p(\bmu,\bSigma,\nu)$ then $\bz=\A\x+\bb\sim t_p(\A\bmu+\bb,\A\bSigma\A^T,\nu)$ with $\A\neq \0$.

Finally, let us consider again the decomposition given in equations (\ref{eq:dec1})--(\ref{eq:dec2}). Arellano-Valle et al. (2013) \cite{arellano2013shannon} provided an expression for the mutual information between $\x_1$ and $\x_2$
\begin{align}
\label{eq:It}
I_t(\x_1,\x_2)=&I_N(\x_1,\x_2)+\zeta(\nu)\quad\mbox{with}\\[5mm]
\zeta(\nu)=&\log\left[\dfrac{\Gamma(\nu/2)\Gamma[(\nu+p_1+p_2)/2]}{\Gamma[(\nu+p_1)/2]\Gamma[(\nu+p_2)/2]}\right]+\dfrac{\nu+p_2}{2}\psi\left(\dfrac{\nu+p_2}{2}\right)+
\dfrac{\nu+p_1}{2}\psi\left(\dfrac{\nu+p_1}{2}\right)-\nonumber\\[8mm]
&\dfrac{\nu+p_1+p_2}{2}\psi\left(\dfrac{\nu+p_1+p_2}{2}\right)-\dfrac{\nu}{2}\psi\left(\dfrac{\nu}{2}\right)\nonumber
\end{align}
where $\psi(x)=d/dx[\Gamma(x)]$ is the so-called digamma function and 
$$
I_N(\x_1,\x_2)=-\dfrac{1}{2}\log\left(\dfrac{\left|\bSigma\right|}{\left|\bSigma_{11}\right|\left|\bSigma_{22}\right|}\right)
$$
Notice that in equation (\ref{eq:It}) $I_N(\x_1,\x_2)$ actually quantifies the mutual information between $\x_1$ and $\x_2$ under the normal model, which can be easily shown by considering $\Var(\x)=\nu /(\nu -2)\bSigma$ (for $\nu >2$). 

Interestingly, all the information due to $\bSigma$ arises only from $I_N(\x_1,\x_2)$ while the information due to $\nu$ comes from the remaining terms. It can also be shown that as $\nu$ increases, the t mutual information converges to the normal mutual information.

\subsection{The t-causal Model}
\label{sec:tcausal-model}

\begin{figure}[!ht]
\begin{center}
\includegraphics[scale=0.6]{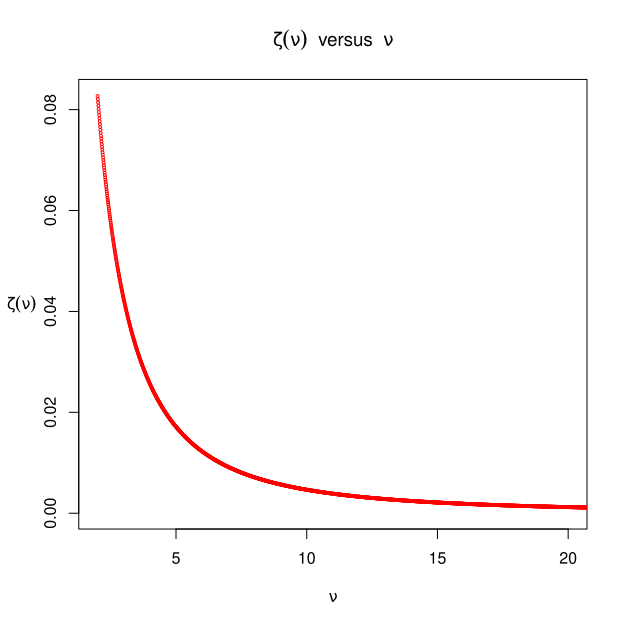}
\end{center}
\caption{$\zeta(\nu)$ as a function of $\nu$}
\label{fig:Ztanu}
\end{figure}

Let us assume that $\Y\sim t_p(\bmu,\bSigma,\nu)$ with $\bmu$ and $\bSigma$ as before. The results presented in Section \ref{sec:MtD} imply that $\bDelta\sim t_p(\bmu_{\Delta},\bSigma_{\Delta},\nu)$. One could now assess Definition \ref{def:surrogate} using the SICC as given in equation (\ref{eq:R2H}), i.e., by working with the expression
$$
R_{Ht}^2=1-e^{-2I_t(\Delta T, \Delta S)}=1-e^{-2I_N(\Delta T, \Delta S)-2\zeta(\nu)}=1-(1-\rho_{\Delta}^2)e^{-2\zeta(\nu)}
$$
where 
\begin{align}
\zeta(\nu)=2\log\left[\dfrac{\Gamma(\nu/2)}{\Gamma[(1+\nu)/2]}\sqrt{\dfrac{\nu}{2}}\,\right]+\left(1+\nu\right)\psi\left(\dfrac{1+\nu}{2}\right)-\left(1+\nu\right)\psi\left(\dfrac{\nu}{2}\right)-\dfrac{2+\nu}{\nu}.
\end{align}
To derive the final equation for $\zeta(\nu)$, we use the properties of the gamma and digamma functions: $\Gamma(1+z) = z\Gamma(z)$ and $\psi(1+z) = \psi(z) + 1/z$. As illustrated in Figure \ref{fig:Ztanu}, when $\nu$ approaches infinity, $\zeta(\nu)$ converges to zero and $R_H^2 = R_{Ht}^2 = \rho_{\Delta}^2=ICA_t$. This is expected, as the multivariate t-distribution converges to a multivariate normal distribution as the degrees of freedom $\nu$ increase. 

\begin{figure}[!t]
\begin{center}
\includegraphics[scale=0.3]{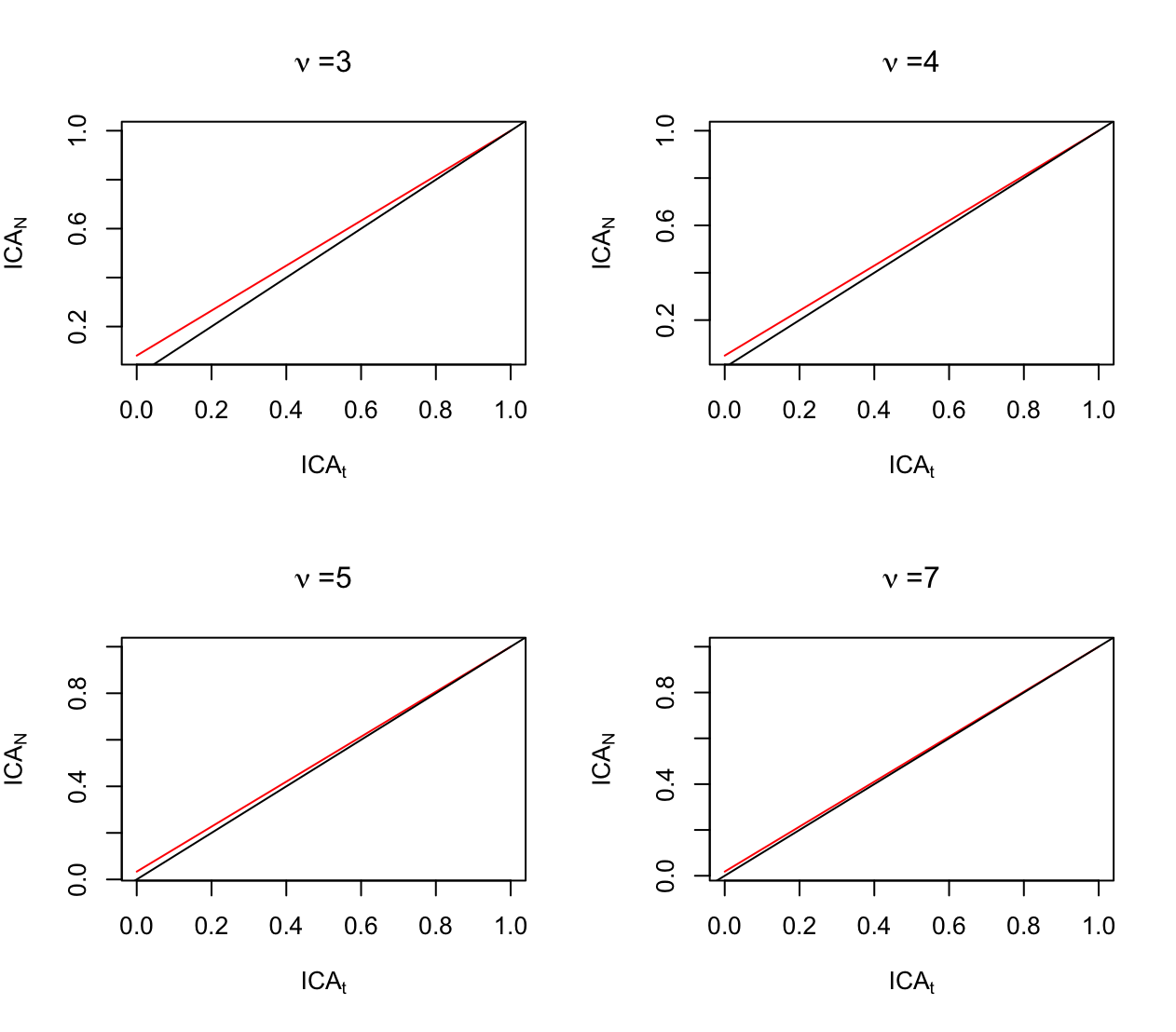}
\end{center}
\caption{$ICA_t$ versus $ICA_N$. The black line is the identity line $y=x$}

\label{fig:RHtvsRH}
\end{figure}

Additionally, Figure \ref{fig:RHtvsRH} plots the pairs $(ICA_t, ICA_N)$ (dashed line) for $(\nu = 3, 4, 5, 7)$, with the continuous line representing the identity function $y=x$ for reference. The figure clearly shows that using the normal causal inference model, when the t-distributed causal model holds, will only have a mild impact on the ICA. Indeed, the effect of the misspecification is noticeable only when $ICA_t$ is small and $\nu = 3$. In this scenario, the normal causal model slightly overestimates the true value of $ICA_t$. When $\nu \geq 4$, the effect of the misspecification becomes negligible. This discussion demonstrates that the normal causal model yields practically meaningful results when the t-causal model is the true underlying data-generating mechanism. 

Although encouraging, this finding should be taken with some caution. The t-distribution shares many properties with the normal distribution and converges to the normal rather quickly as the degrees of freedom increase. Therefore, the next section will explore the effect of misspecification using a log-normal distribution. Addressing this case purely from a mathematical standpoint is not feasible due to the complexity of the algebra involved. Consequently, a Monte Carlo procedure will be employed. First, however, we will present an extension of the algorithm introduced by Alonso et al. (2015) \citep{alonso2015relationship} to assess the ICA based on the normal causal model for scenarios where other models are used.

\section{Sensitivity Analysis Algorithm}
\label{sec:ext-alg}
Under the normal causal inference model, the vector of correlations $\btheta$ cannot be estimated from the data, rendering $\rho_{\Delta}$ unidentifiable. To address this issue, Alonso et al. (2015) \cite{alonso2015relationship} proposed a simulation-based sensitivity analysis. However, their approach is limited to the normal case. Therefore, we opted to develop a more versatile algorithm that could be applied to a broader range of scenarios. To that end, let us consider the causal model $\Y\sim F(\y|\btheta)$ with  $F$ a four-dimensional distribution function. Let us now partition the vector of parameters $\btheta=(\btheta_I,\btheta_U)$ with $\btheta_I$ and $\btheta_U$ denoting the parameters that are identifiable and unidentifiable, respectively. Furthermore, let $\hat{\btheta}_I$ denote the parameter estimates of $\btheta_I$. Note that, in general, $\hat{\btheta}_I$ may depend on $\btheta_U$. In other words, once a value is assigned to the unidentifiable parameters, $\btheta_I$ can be estimated from the data. Again, the ICA can be conceptualized as a mathematical function of $\btheta_U$; specifically, we aim to study the behavior of $R_{H}^2(\btheta_U)$. To assess the ICA, the following algorithm can be employed:
\begin{enumerate}
    \item Sample $\btheta_U$ on $\Gamma_D=\{\btheta_U:\mbox{ so that }F\left(\y|\hat{\btheta}_I,\btheta_U\right)\mbox{ is a valid distribution}\}$\label{step1a}
    \item Generate $M_y$ vectors of potential outcomes $\Y = (T_0, T_1, S_0, S_1)^T$ using $\Y \sim F(\y | \hat{\btheta}_I, \btheta_U)$ based on the values of $\btheta_U$ obtained in step \ref{step1a}. The value of $M_y$ should be sufficiently large to ensure an accurate approximation of the mutual information in subsequent steps. For instance, consider $M_y = 2000$ or $3000$, depending on the available computing resources.\label{step2a}
    \item Using the ($M_y$) $\Y$ vectors obtained in step \ref{step2a}, calculate the vectors of individual causal treatment effects $\bDelta=(\Delta T, \Delta S)^T$.\label{step3a}
    \item Based on the vectors $\bDelta$ obtained in the previous step, estimate the mutual information between the individual causal treatment effects using, for instance, the \texttt{mutinfo()} function in the \texttt{FNN} package (2024) in R. Finally, estimate the ICA as given in equation (\ref{eq:R2H}). \label{step4a}
    \item Repeat steps \ref{step1a}--\ref{step4a} $N$ times. \label{step35}
\end{enumerate}
The algorithm will generate $N$ values for the ICA, and their frequency distribution can be analyzed to understand the behavior of $R_{H}^2(\btheta_U)$ on $\Gamma_D$. In each iteration of the algorithm, $M_y$ vectors $\bDelta$ are used to estimate the mutual information between the individual causal treatment effects and, hence, the ICA. The larger the $M_y$, the more precise our estimate of the ICA will be at each iteration.

For certain distributions $F$ the $M_y$ vectors in step \ref{step1a} can be directly generated like, for example, when $F$ is a multivariate normal, t or a log-normal distribution. However, in other scenarios one may need to resort to more general Markov chain Monte Carlo (MCMC) algorithms like, for instance, the Metropolis Hastings algorithm to implement the generation process. This may increase the computational burden but, at the same time, it may allow the use of very flexible models to describe the vector of potential outcomes $\Y$.

\section{The Log-normal Causal Model}
\label{sec:lognormal}
In Section \ref{sec:multi_t_dist}, we showed that assuming a normal causal inference model, when the true model is based on a multivariate t-distribution, generally has a negligible impact on the ICA value that would be estimated. However, exploring this issue for other types of multivariate distributions, such as the multivariate log-normal distribution, becomes mathematically challenging due to the lack of close form expressions. Therefore, to dive deeper into this problem, we will resort to a Monte Carlo procedure in this section. To that end, we now assume that the vector of potential outcomes $\Y$ follows a four-dimensional log-normal distribution with the density function
\begin{equation*}
f(\y|\bmu_y, \bSigma)= \dfrac{1}{\sqrt{2\pi|\bSigma|^4}\prod_{i=1}^4 y_i}\, e^{-\dfrac{\left(\ln(\y)-\bmu\right)^T\bSigma^{-1}\left(\ln(\y)-\bmu\right)}{2}}.
\end{equation*}
For the multivariate log-normal causal model, the distribution of the individual causal treatment effects $\bDelta$ does not have a closed form, and hence, the ICA cannot be computed analytically. To approximate the distribution of $\bDelta$ and calculate the corresponding ICA, we used Monte Carlo simulations. Specifically, we generated 200 different pairs of $\bmu$ and $\bSigma$ for the underlying log-normal distribution, using a normal distribution for $\bmu$ and a Wishart distribution for $\bSigma$. For each pair (setting), we generated $M_y = 2000$ vectors of potential outcomes $\Y$. The true ICA value (as given in equation \ref{eq:R2H}) for each of these settings was approximated by applying steps \ref{step3a} and \ref{step4a} of the algorithm introduced in Section \ref{sec:ext-alg} to the previously generated $2000$ vectors $\Y$. This true ICA value will be denoted as $ICA_L$.

Additionally, we also computed the ICA under the assumption that the vector of potential outcomes $\Y$ follows a multivariate normal distribution, with the same mean and variance as the correct log-normal distribution. This was done by using the $M_y = 2000$ vectors of potential outcomes $\Y$ generated in each setting to obtain the individual causal treatment effects vectors $\bDelta$ and calculate $R_H^2 = \rho_{\Delta}^2=\mbox{corr}(\Delta T,\Delta S)^2$. We denote the ICA calculated under the normal assumption as $ICA_N$. 

\begin{figure}
 \centerline{\includegraphics[width=6in]{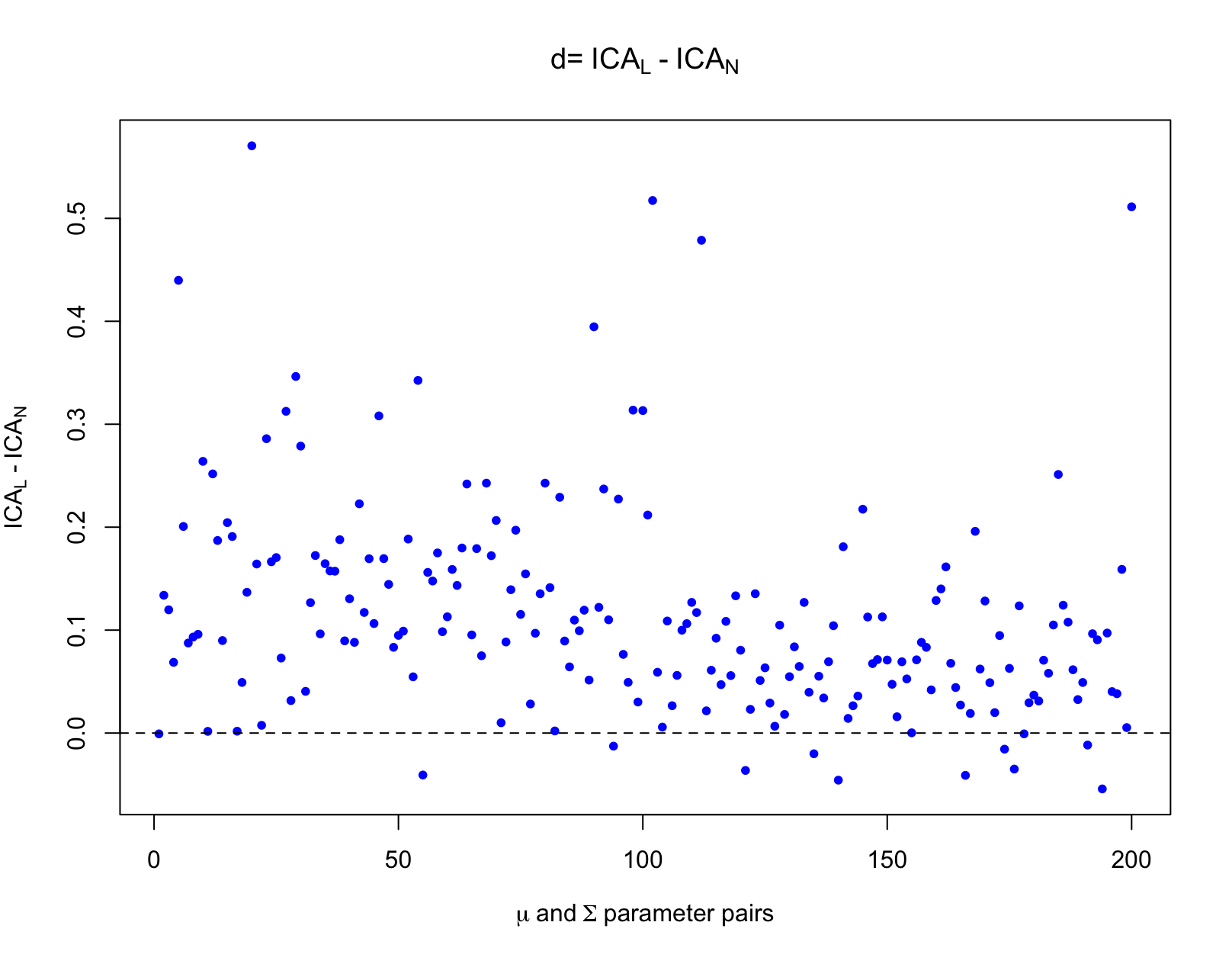}}
\caption{Difference between $ICA_L$ and $ICA_N$}
\label{f:diff_lognormal_and_normal}
\end{figure}

The main findings are summarized in Figure \ref{f:diff_lognormal_and_normal}, which displays the values of the difference $d = ICA_L - ICA_N$. It is evident from the figure that using a misspecified model can significantly impact the results in some cases. The maximum observed value for $d$ was $0.571$, indicating a substantially smaller ICA under the normal model. Although the misspecified model generally yields smaller ICAs than the correct model ($d > 0$), it can also produce larger values, with the minimum difference observed being $d = -0.0543$.

We further explored the settings where this difference was small and large, as defined in Web Appendix A. We observed that in cases where the difference between $ICA_L$ and $ICA_N$ was substantial, say larger than $0.3$, the underlying log-normal distribution used to generate the potential outcomes was notably different from a normal distribution. This includes the distribution of the identifiable margins, suggesting that one will likely be able to detect the misspecification in those cases.

\section{Case Studies}
\label{sec:copula}

In practice, the true data-generating mechanism is unknown, and some parts of the causal inference model are untestable. One way to address this problem is to consider several models that fit the observed data equally well and compare the results they deliver. Hence, the model becomes an integral part of the sensitivity analysis. In this section, we implement this approach, using D-vine copulas, in the analysis of two real-life data sets:

\begin{enumerate}
    \item The age-related macular degeneration (\textit{ARMD}) data set contains data from a randomized clinical trial in ARMD where the change in visual acuity at 24 weeks is a potential surrogate for the change in visual acuity at 52 weeks \citep{1997interferon}.
    \item The schizophrenia (\textit{Schizo}) data set combines data from five randomized clinical trials in schizophrenia, considering the positive and negative syndrome scale (PANSS)  as a potential surrogate for the clinical global impression (CGI) scale.
\end{enumerate}
These data sets are available in the \textit{Surrogate} R package (2024) \citep{van2023package}, and we refer interested readers to the package documentation for further details.

\subsection{D-vine Copula Model}

The objective of this subsection is to develop a model that is indistinguishable, based on the observed data, from the multivariate normal model in Section \ref{sec:Normal}. We begin by introducing copulas and D-vine copulas. Next, we develop the required model using a D-vine copula.

Copulas, further denoted by $C: [0, 1]^d \to [0, 1]$, are $d$-dimensional distribution functions with uniform margins. They are useful in applied statistics because they allow us to describe the association between random variables independently of their marginal distributions \citep{sklar1959fonctions}.
A copula $C$, which is a distribution function, has a corresponding \textit{copula density} $c$ that is obtained by partial differentiation; for instance, for a bivariate copula we have that $c(u, v) = \frac{\partial^2}{\partial u \partial v} C(u, v)$ is the copula density. 
The best known and simplest parametric copulas are bivariate; furthermore, $d$-dimensional copulas can be constructed using only bivariate copulas as building blocks, based on the D-vine copula construction \citep{czado2019analyzing}. Specifically, the corresponding D-vine copula density is the product of a particular set of conditional and unconditional copula densities (see next paragraph). 
A \textit{conditional} copula density (e.g., $c_{T_0 S_1;S_0}$) is simply the copula density corresponding to a conditional distribution (e.g., $(T_0, S_1)^T \mid S_0$).  
Further details on (D-vine) copulas and related concepts are provided Web Appendix B.1. 

Let $f_{\boldsymbol{Y}}$ be the joint density of $\boldsymbol{Y} = (T_0, S_0, S_1, T_1)^T$ (note the reordering). The D-vine density decomposition of $f_{\boldsymbol{Y}}$ is the product of four marginal densities and six bivariate copula densities:
\begin{equation}\label{eq:d-vine-model-density}
    f_{\boldsymbol{Y}} =  f_{T_0} \, f_{S_0} \, f_{S_1} \, f_{T_1} 
             \cdot c_{T_0 S_0} \, c_{S_0 S_1} \, c_{S_1 T_1} 
             \cdot c_{T_0 S_0;S_1} \, c_{S_0 T_1;S_1} 
             \cdot c_{T_0 T_1;S_0 S_1},
\end{equation}
where (i) $f_{T_0}$, $f_{S_0}$, $f_{S_1}$, and $f_{T_1}$ are univariate density functions, (ii) $c_{T_0, S_0}$, $c_{S_0, S_1}$, and $c_{S_1, T_1}$ are unconditional bivariate copula densities, and (iii) $c_{T_0, S_1; S_0}$, $c_{S_0, T_1; S_1}$, and $c_{T_0, T_1; S_0, S_1}$ are conditional bivariate copula densities.
A conditional copula density can depend on the conditioning variable in arbitrary ways as long as it corresponds to a valid copula density for any fixed value of the conditioning variable. Consequently, (\ref{eq:d-vine-model-density}) is intractable to use for constructing models. The simplifying assumption (Definition B.3 in the Web Appendix) is, therefore, commonly made in practice \citep{czado2019analyzing}. This assumption implies that the three conditional copula densities in (\ref{eq:d-vine-model-density}) do not depend on the value of the conditioning variable; this greatly simplifies modeling.

The observable bivariate margins follow immediately from the components in (\ref{eq:d-vine-model-density}) as follows:
\begin{equation*}
    f_{S_0 T_0}(s, t) = f_{S_0}(s) f_{T_0}(t) c_{T_0 S_0} \left\{ F_{T_0}(t), F_{S_0}(s) \right\} \text{ and } f_{S_1 T_1}(s, t) = f_{S_1}(s) f_{T_1}(t) c_{S_1 T_1} \left\{ F_{S_1}(s), F_{T_1}(t) \right\}.
\end{equation*}
If the marginal distribution functions are normal, and $c_{T_0 S_0}$ and $c_{S_1 T_1}$ are Gaussian copulas, then $f_{S_0 T_0}$ and $f_{S_1 T_1}$ are bivariate normal densities. Hence, the distribution in (\ref{eq:d-vine-model-density}) is then indistinguishable from the multivariate normal model \textit{regardless of the parametric choices for} $c_{T_0, S_1; S_0}$, $c_{S_0, T_1; S_1}$, and $c_{T_0, T_1; S_0, S_1}$ because the latter copula densities do not affect $f_{S_0 T_0}$ and $f_{S_1 T_1}$.

\subsection{Analysis of the data}

In this subsection, we empirically investigate the effect of unverifiable parametric assumptions on the ICA using D-vine copula models that fit the observed data equally well. (Details in Web Appendix B.2)

In our investigations, we fix the two observable bivariate margins, $f_{S_0 T_0}$ and $f_{S_1 T_1}$, at the estimated bivariate normal distributions for $(S_0, T_0)^T$ and $(S_1, T_1)^T$ and consider the following four parametric copulas for the four unidentifiable copulas in (\ref{eq:d-vine-model-density}): (i) Gaussian, (ii) Clayton, (iii) Gumbel, and (iv) Frank.
We consider all combinations of these parametric copulas, leading to $4^4$ different D-vine copula models where four times Gaussian leads to the multivariate normal model. Regardless of the chosen parametric copulas, the four unidentifiable copulas can be parameterized by Spearman's rho correlation parameters. Along the lines of Alonso et al. (2015) \cite{alonso2015relationship} and the ideas presented in Section \ref{sec:ext-alg}, we sample $1000$ sets of four Spearman's rho parameters (one for each unverifiable copula). This is repeated under three sampling schemes:
\begin{itemize}
    \item \textit{No additional assumptions.} The rho parameters are sampled from $U(-1, 1)$.
    \item \textit{Positive restricted associations.} The rho parameters are assumed to be positive and bounded away from zero and one. 
    \item \textit{Conditional independence and positive restricted associations.} In addition to positive restricted associations, we assume conditional independence: $T_0 \perp S_1 \mid S_0$ and $T_1 \perp S_0 \mid S_1$.
\end{itemize}
For all combinations of (i) data set, (ii) sampling scheme, (iii) sampled rho parameters, and (iv) D-vine copula model, we compute the ICA. This leads to $4^4$ computed ICAs per set of sampled rho parameters in a given sampling scheme and data set; corresponding to the $4^4$ D-vine copula models. 
We analyze the impact of the unverifiable part of the model on the ICA by pairing the $ICA_{N}$, computed under the multivariate normal model, with the ICAs under the remaining $4^4 - 1$ D-vine copula models under the same (i) data set, (ii) sampling scheme, and (iii) sampled rho parameters. The only difference between such paired ICAs are the untestable parametric assumptions. 
Further technical details about this approach are given in Web Appendix B. 

\begin{figure}
	\centering
	\includegraphics[width=0.7\linewidth]{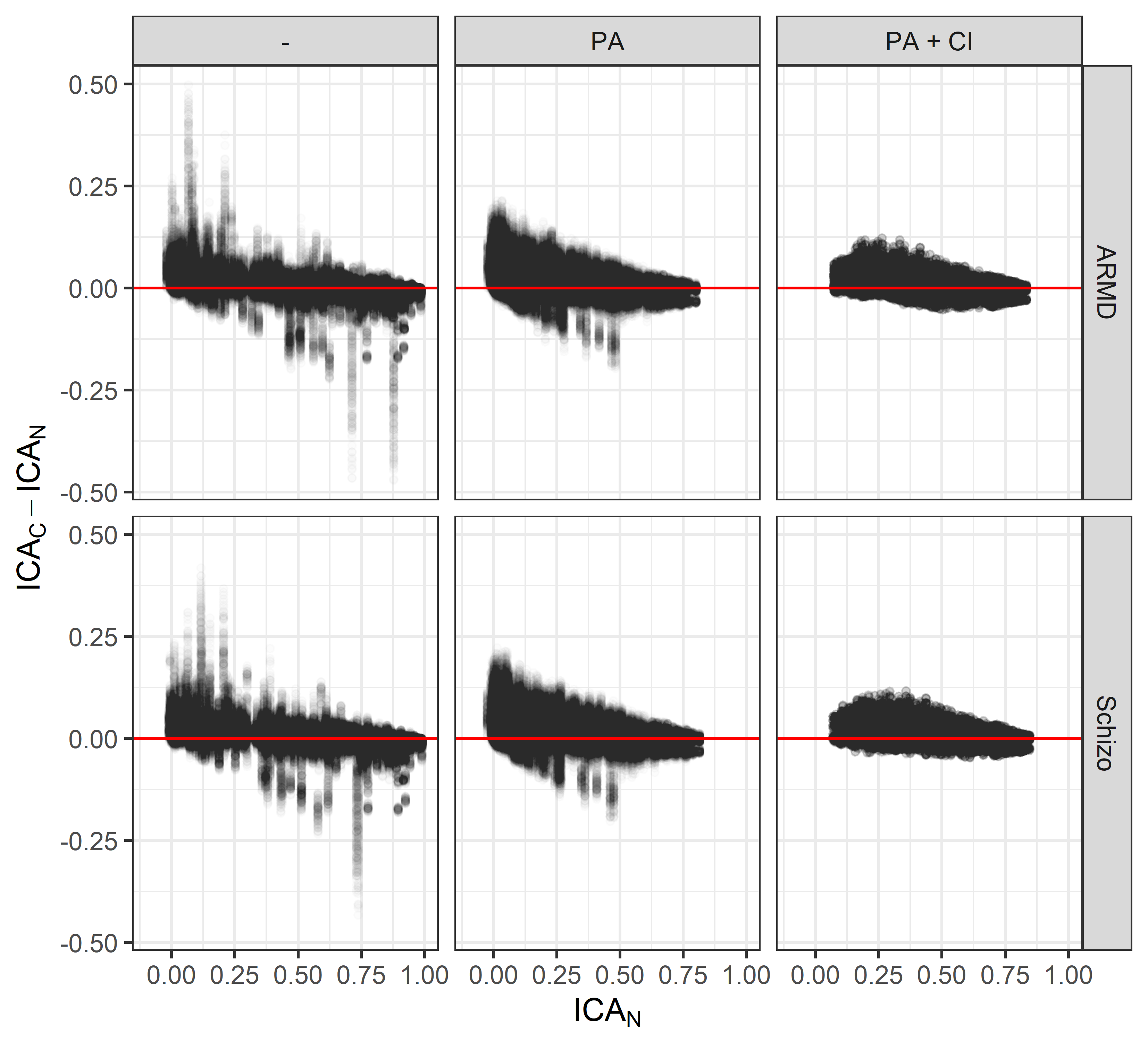}
	\caption{Results of the analysis of the ARMD and Schizo data sets under various D-vine copula models versus the multivariate normal model. The columns correspond to the additional sets of assumptions, the rows correspond to the data sets.
    $ICA_C - ICA_N$: difference between the computed ICA under a D-vine copula model ($ICA_C$) and the ICA under the corresponding multivariate normal model ($ICA_{N}$). -: No additional assumptions; PA: Positive restricted associations; CI: conditional independence.}
	\label{fig:vine-copula-mvn-reference}
\end{figure}

All ICAs are plotted in Figure \ref{fig:vine-copula-mvn-reference}, with the x-axis representing the $ICA_{N}$ and the y-axis representing $ICA_C - ICA_N$ where $ICA_C$ is an ICA value paired to $ICA_N$. This difference shows the impact of the unverifiable parametric assumptions in the D-vine copula models.
When no additional assumptions are made for Spearman's rho (first column), the ICAs under some D-vine copula models differ substantially from those computed under normality.
Under the positive restricted associations assumption (second column), the differences are smaller and decrease as the ICA increases.
Finally, under the assumption of conditional independence and positive restricted associations (third column), the differences are close to zero, indicating a minor impact of the unverifiable parametric assumptions. 

The D-vine copula models used in the above analyses may produce different results, depending on which additional assumptions are made, even though they fit the observed data equally well.

\section{Discussion}
\label{sec:discuss}

In this work, we investigated the effects of model misspecification on the behavior of ICA when the true distribution deviates from the multivariate normal model through both theoretical and numerical studies. Specifically, we evaluated it theoretically when the potential outcome vector follows a multivariate t-distribution and through numerical studies when the model follows a multivariate log-normal distribution. Additionally, we assessed the impact of unverifiable assumptions on the assessment of surrogacy using D-vine copula models that are indistinguishable from the multivariate normal model in terms of observable data.

Our exploration demonstrates that the impact of model misspecification can range from negligible (e.g., when the underlying model is based on a multivariate t-distribution) to substantial (e.g., when the underlying model is based on a multivariate log-normal distribution). However, in the latter case, the impact was significant only when the deviation from normality was considerable.

In our case study analysis, we demonstrated that models fitting the observable data equally well can still yield different ICA values depending on the assumptions made for the unidentifiable parameters. In such situations, the normal model could still be justified as a reference point (if it describes the observed data) by invoking the maximum entropy principle (MEP) \citep{Alonso2019b}. The MEP suggests that, when making inferences based on incomplete information, one should select the probability distribution with the highest entropy given the known constraints. Among all continuous distributions with a specified mean and variance, the normal distribution maximizes entropy, making it the most ``uninformative'' or ``least biased'' choice under these conditions. However, it is advisable to interpret the results within the framework of a sensitivity analysis that considers alternative models, ensuring that conclusions are robust across different modeling assumptions.

\bmsection*{Acknowledgments}
This work is supported by grant PID2022-137050NB-I00 of the Spanish Ministry of Science and Innovation.\\
Florian Stijven gratefully acknowledges funding from Agentschap Innoveren \& Ondernemen and Janssen Pharmaceutical Companies of Johnson \& Johnson Innovative Medicine through a Baekeland Mandate [grant number HBC.2022.0145].

\bmsection*{Conflict of interest}

The authors declare no potential conflict of interests.

\bibliography{wileyNJD-AMA}

\bmsection*{Supporting information}

Additional supporting information may be found in the
online version of the article at the publisher’s website.

\end{document}